\documentclass[preprint,11pt]{JHEP3} 

\JHEPspecialurl{http://jhep.sissa.it/JOURNAL/JHEP3.tar.gz}

\usepackage{epsfig,multicol,amsmath,slashed}
\usepackage[square, comma, sort&compress]{natbib}

\def\beq{\begin{equation}}
\def\eeq{\end{equation}}
\def\bea{\begin{eqnarray}}
\def\eea{\end{eqnarray}}

\def\eq#1{{Eq.~(\ref{#1})}}
\def\fig#1{{Fig.~\ref{#1}}}
\def\sec#1{{section~\ref{#1}}}

\parskip=\itemsep               

\newcommand{\nn}{\nonumber}

\newcommand{\h}{\frac{1}{2}}

\newcommand{\al}{\alpha}

\newcommand{\de}{\delta}

\newcommand{\ka}{\kappa}

\newcommand{\Ga}{\Gamma}

\newcommand{\f}{\frac}

\newcommand{\If}{\int^{\infty}_{-\infty}}

\newcommand{\bas}{\bar{\alpha}_s}
\newcommand{\as}{\alpha_s}

\newcommand{\Lb}{\left(}
\newcommand{\Rb}{\right)}

\newcommand{\intf}{\int^\infty_{-\infty}}

\newcommand{\tom}{\widetilde{\omega}}
\newcommand{\lab}{\label}

\title{\LARGE \bf The Schwinger-Dyson equation on Pomeron loop summation and renormalization}

\author{\large  J.~Miller\thanks{Email:
jeremy.miller@ist.utl.pt; miller@physics.org}\,\, \\
CENTRA, Departamento de F$\acute{i}$sica, Instituto Superior T$\acute{e}$cnico (IST),\\  Av. Rovisco Pais,\\1049-001 Lisboa,\\Portugal}

\abstract{The solution to the Schwinger Dyson equation that describes the summation over Pomeron loop diagrams is derived. The solution is
a closed expression which splits into two parts. The first leads directly to the renormalization of the BFKL Pomeron, and the second contribution
is equivalent to non interacting Pomerons with renormalized vertices. Thus a closed expression is derived for the sum over Pomeron loop diagrams
in the perturbative QCD approach, which
preserves unitarity.}

\preprint{ \today}

\begin{document}

\section{Introduction}

\FIGURE[h]{\begin{minipage}{70mm}
\centerline{\epsfig{file=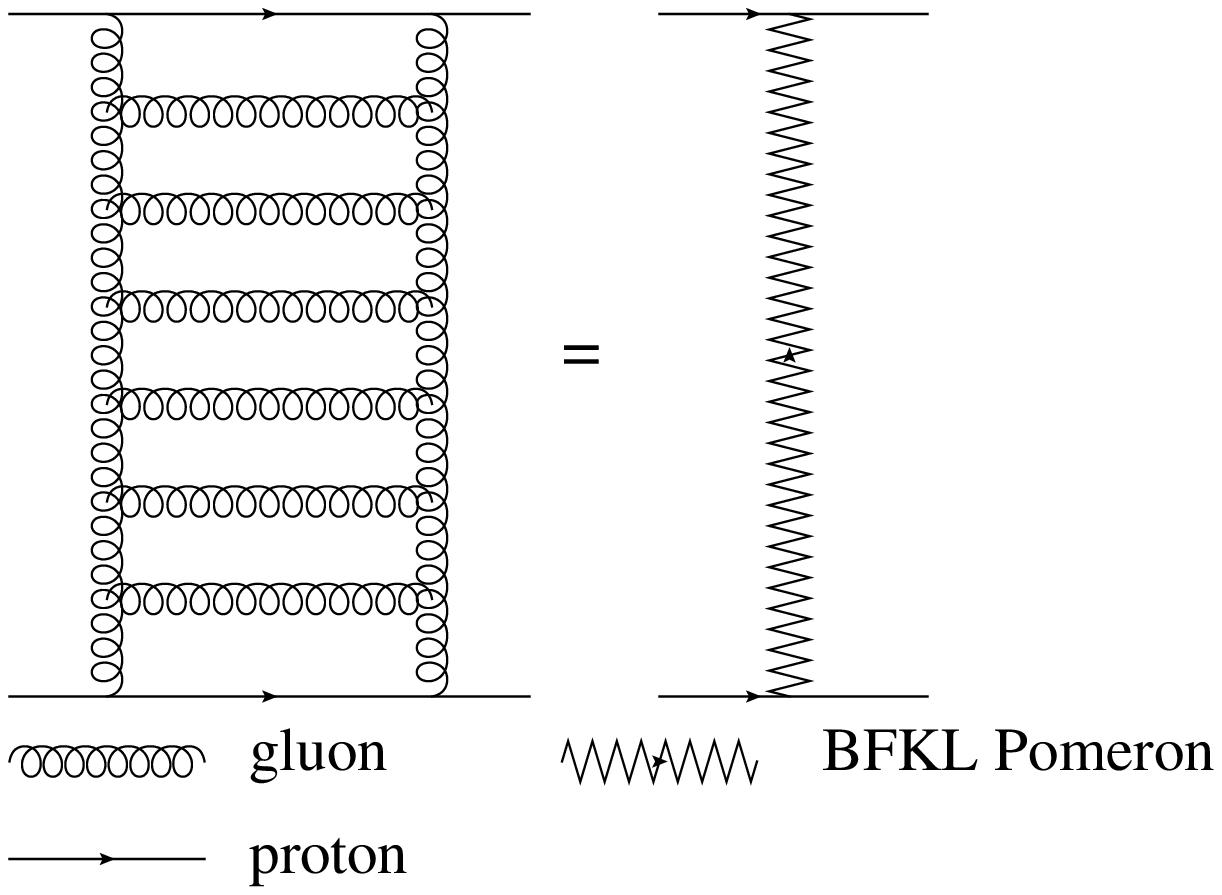,width=70mm}}
\end{minipage}
\caption{ The BFKL Pomeron} \label{fBFKL} }

The goal of this paper is to derive a solution to the Schwinger-Dyson equation, which provides the full sum over Pomeron loop diagrams.
The motivation for pursuing the summation over Pomeron loop diagrams, is the large contribution of Pomeron loops to the diffractive scattering amplitude in short distance interactions.
Hence a reliable calculation of the scattering amplitude demands the summation over Pomeron loop diagrams to be taken into account. The bare scattering amplitude from the t-channel
exchange of a single Pomeron, and also Pomeron loop diagrams grow with energy. Unitarity is only restored by replacing the Pomeron Green's function with the
sum over the full set of loops. \\

The BFKL Pomeron shown in \fig{fBFKL} is the t-channel exchange of a pair of  vertical gluons, interacting through horizontal gluons which form the ``rungs of the ladder" structure.
The BFKL Pomeron is the sum over all ladder diagrams of this type with $n$ rungs of the ladder.
The vertical gluons are themselves a superposition of the sum over $n$ rung ladder diagrams, and so on. This
leads to the scattering amplitude $A\Lb s,t\Rb$ which was derived in refs. \cite{Gribov:1983,
Bartels:1975,Lipatov:1989,Cheng:1976,Ross} in the leading log approximation to be proportional to

\beq
A\Lb s,t\Rb\propto\sum^\infty_{n=0}\f{1}{n!}\Lb \al_G\Lb t\Rb \ln\Lb s/s_0\Rb\Rb^n=s^{\al_G\Lb t\Rb}\lab{LLA}\eeq

 where $\al_G\Lb t\Rb$ is the Regge trajectory.
In this way the sum over ladder diagrams is achieved by replacing the two interacting vertical gluons with a ``reggeon" which behaves as $s^{\al_G\Lb t\Rb}$ at high energy. According to the optical theorem
the  total cross section  behaves as $\sigma_{tot}\propto s^{\al_G\Lb 0\Rb-1}$.  Experimentally it is known that the total cross section rises slowly with $s$ which means
 $\al_G\Lb 0\Rb >1$.  Pomeranchuk
 first commented \cite{Pomeranchuk:1956,Okun:1956} that this behavior is matched by the theoretical prediction that $\al_G\Lb 0\Rb>1$ when the t-channel exchange carries zero quantum numbers,
 including zero charge and color flow. Such particles with quantum numbers of the vacuum exist in QCD
 for bound gluon states. This kind of bound state trajectory is called the Pomeron named after Pomeranchuk, which is the double t-channel gluon exchange shown in \fig{fBFKL}.
The evolution of the vertical t-channel gluons  to the sum over ladder diagrams is called the BFKL Pomeron
which is described by the BFKL equation \cite{Fadin:1976,Balitsky:1978}. \\

For hard collisions,
Pomeron loop corrections contribute substantially to the scattering amplitude, and a reliable estimate of the scattering amplitude requires the sum over Pomeron loop diagrams.
This is a difficult problem, and for a long time the generally accepted method for estimating the sum over Pomeron loop diagrams was
the
Mueller, Patel, Salam and Iancu (MPSI) approach,
(see refs. \cite{Mueller:1996te,
Salam:1995uy,Iancu:2003uh,Iancu:2003zr,
Levin:2007yv,toy,
Levin:2007wc}).  A. Mueller \cite{toy} and Levin. et. al. \cite{Levin:2007wc} first commented that at high
energy Pomeron loop diagrams should reduce to independent Pomeron exchanges with complicated non factorized vertices.
In refs. \cite{Miller:2009ca,Miller:2009} it was shown that this problem can be solved theoretically in perturbative QCD for a specific type of diagrams. In this approach the special class of symmetric diagrams of the type shown in \fig{fsopl} were calculated
using an iterative technique, based on the observation that \fig{fsopl} (b) is generated from \fig{fsopl} (a) when each branch of the loop gives birth to a secondary loop leading to two  ``second generation" loops. Likewise
\fig{fsopl} (c) arises when when each of the two ``second generation"  loops  in \fig{fsopl} (b) gives birth to two loops which leads to four ``third generation" of loops. Hence the diagrams in \fig{fsopl} are called the $N=1$, $N=2$ and $N=3$
generation loop diagrams, and continuing in this way one can generate the full spectrum of symmetric Pomeron loop diagrams, with $N$ generations of loops. \\

The formula derived in ref.  \cite{Miller:2009ca,Miller:2009}  for the sum over this class of symmetric diagrams was of the type $\sum_N A\Lb N\Rb$, where $ A\Lb N\Rb$ is the diagram with $N$ generations of loops.
This formula suffered from the
difficulty that it diverges with energy. In this paper, this problem  of unitarity violation is resolved by taking instead the sum $\sum_n A\Lb n\Rb$ where $A\Lb n\Rb$ is the symmetric
diagram with $n$ Pomeron loops.
 In this way a closed analytic expression is obtained, which is equivalent
to the sum over diagrams with $2n$ non interacting Pomerons.\\

However this does not complete the sum over Pomeron loops, since the class of diagrams with successive loops shown in \fig{fren} should also be included.
It was first suggested by M. Braun in refs. \cite{Braun:2005hx,Braun:2009sh} that the Schwinger-Dyson equation automatically generates the sum over the full spectrum
of Pomeron loop diagrams, which includes both types of diagrams shown in \fig{fren} and \fig{fsopl}.\\

This paper is organized in the following way. Firstly in \sec{s1P} the scattering amplitude arising from a t-channel bare Pomeron shown in \fig{fBFKL} is derived, for the sake of completeness.
 Next a solution to the Schwinger-Dyson equation is presented which splits into parts.
The first part discussed in \sec{sSD} leads directly to a simple expression for the renormalized Pomeron intercept, which arises from summing over the class of diagrams in \fig{fren}. The second part of the solution
derived in \sec{sNIP}
is equivalent to the sum over non interacting Pomeron diagrams, which is derived from the sum over the type of diagrams in \fig{fsopl}. Intuitively this can be seen from an observation of \fig{fsopl}, where at high energy taking the
branches of the loop in \fig{fsopl} (a) outside leads to 2 non interacting Pomerons. Likewise for the 2 small loops in \fig{fsopl} (b) that have not given birth to any more smaller loops, taking
the branches of the loops outside leads to 4 non interacting Pomerons. In \sec{sR} the main results of the paper are presented, and
in \sec{cad} the conclusions of this paper are discussed.

\section{The bare Pomeron scattering amplitude}
\label{s1P}

In this section the diffractive  scattering amplitude arising from a t-channel bare Pomeron shown in \fig{fBFKL}, is derived.
Although the expression is well known and has been calculated in refs. \cite{Miller:2006bi,Levin:2007wc,Miller:2009ca,Miller:2009,Kozlov:2004sh,Navelet:2002zz,Navelet:1998yv,Navelet:1997xn},
it has been included for the sake of completeness.
The solution to the BFKL equation
provides the trajectory $\omega\Lb n,\nu\Rb$ $(n\in \mathbb{Z}\,;\,\nu\in\mathbb{R})$ for the BFKL Pomeron as \cite{Fadin:1976,Balitsky:1978};

\bea
\omega\Lb n,\nu\Rb=\bas\Lb\psi\Lb 1\Rb-\Re e\,\psi\Lb \f{1+n}{2}+i\nu\Rb\Rb;\hspace{1cm}\bas\equiv\f{\as N_c}{\pi}\label{BFKLeigenvalue}\eea

where $\psi\Lb x\Rb=d\ln\Ga\Lb x\Rb/dx$ is the di-gamma function, $n$ represents the energy levels of the BFKL Pomeron, and $\nu$ is a continuous variable which one integrates over when calculating Feynman diagrams. The BFKL eigenfunction falls sharply with increasing $n$ and
is only positive at high energy when $n=0$. Hence throughout this paper which is focussed on high energy scattering, $n=0$ is assumed and the argument $n$ is suppressed.  Hence the BFKL Pomeron trajectory
which is the sum over ladder diagrams of the type shown in \fig{fBFKL} is described by the regge behavior $s^{\omega\Lb \nu\Rb}\equiv e^{\omega\Lb \nu\Rb y}$. The scattering amplitude
of \fig{fBFKL} is given by the expression;

\bea
&&A_{(0)}\Lb y,\de y_H|\mbox{\fig{fBFKL}}\Rb\!=\!\f{\as^2}{4}\!\int^\infty_{-\infty}\!\!\!d\nu h\Lb\nu\Rb g\Lb\nu\Rb e^{\omega\Lb \nu\Rb y}E_\nu E^\prime_{-\nu}\label{A0}\\
&&h\Lb\nu\Rb=\f{2}{\pi^4}\nu^2;\hspace{0.5cm}g\Lb\nu\Rb=\f{1}{16}\f{1}{\Lb 1/4+\nu^2\Rb^2};\hspace{0.5cm}E_\nu=\Lb\f{r_{12}}{r_{10}r_{20}}\Rb^{1/2+i\nu}\Lb\f{r^\ast_{12}}{r^\ast_{10}r^\ast_{20}}\Rb^{1/2-i\nu}\label{hla}\eea

 $h\Lb\nu\Rb$ is the integration measure which preserves conformal invariance \cite{Braun:2009sh,Braun:2005hx}, $g\Lb\nu\Rb$ is the Pomeron propagator in the conformal basis \cite{Braun:2009sh,Braun:2005hx} and $E_\nu$ is the coupling
 of the BFKL Pomeron to the QCD color dipole \cite{Navelet:1997xn,Braun:2009sh,Braun:2005hx}, in the dipole approach to proton proton scattering. Here $r_{12}=r_1-r_2$ is the transverse size of the dipole and $r_{10}=r_1-r_0$
 where $r_0$ is the center of mass coordinate of the dipole. The observation that the BFKL eigenfunction \eq{BFKLeigenvalue} has a saddle point $\nu=0$ means that one can expand
  the exponential in \eq{A0} as

  \bea
 \omega\Lb\nu\Rb=\omega\Lb 0\Rb-\h\nu^2\omega^{\prime\prime}\Lb 0\Rb+\mathcal{O}\Lb\nu^2\Rb;\hspace{0.5cm} \omega\Lb 0\Rb=4\bas\ln 2;\hspace{0.5cm}\omega^{\prime\prime}\Lb 0 \Rb=28\bas\zeta\Lb 3\Rb\,.
 \label{BFKLexpansion}\eea

 where $\zeta\Lb 3\Rb=1.202$ is the Riemann zeta function. Using this expansion the integration in \eq{A0} is evaluated by the steepest descent method which yields the result \cite{Miller:2009ca,Miller:2009};

  \bea
A_{(0)}\Lb y,\de y_H|\mbox{\fig{fBFKL}}\Rb\!&&=\!\f{\bas^2\Lb 2\pi\Rb^{1/2}\,}{2\pi^2N_c^2}\! \f{e^{\omega\Lb 0\Rb y}}{\Lb\omega^{\,\prime\prime}\Lb 0\Rb y\Rb^{3/2}}\label{A01}\eea

\section{The Schwinger-Dyson equation}
\lab{sSD}

\FIGURE[h]{\begin{minipage}{80mm}
\centerline{\epsfig{file=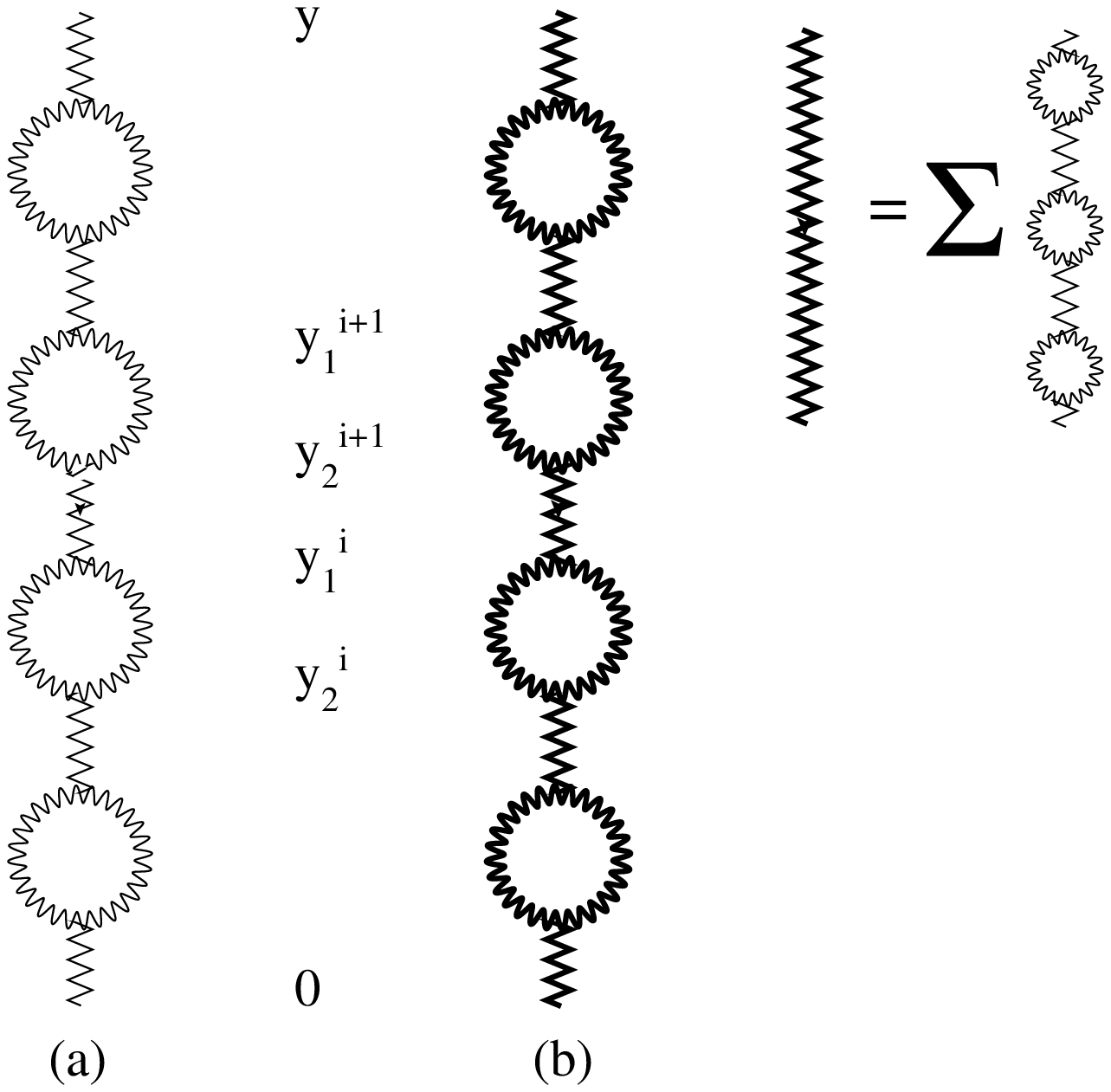,width=80mm}}
\end{minipage}
\caption{Diagram (a) shows $n$ Pomeron loops in series. Diagram (b) shows
a series of $n$ Pomeron self mass interactions, where the bold lines represent a superposition
of  $n$ loops in series.
\vspace{0.1cm}
} \label{fren} }

\eq{A01} is the scattering amplitude arising from the exchange of a bare Pomeron.
According to the Schwinger-Dyson equation \cite{Braun:2005hx}, the bare Pomeron propagator should be replaced by
the Green's function which is found by summing over the class of diagrams of
\fig{fren}.   This Green's function is found by first summing over all diagrams with $n$ consecutive loops in series shown in \fig{fren} (a), from $n=0$ to infinity.
Next, the Pomeron lines themselves in \fig{fren} (a) should be replaced with the sum over loops in series which leads to \fig{fren} (b).
The self mass terms in \fig{fren} (b) are  ``loops made of a superposition of loops". Continuing  in the same way to introduce more generations of loop series, one derives
the Green's function which is the sum over all Pomeron loop diagrams. The process which has been
explained here in words, is described by the Schwinger Dyson equation introduced by M. Braun in ref. \cite{Braun:2005hx,Braun:2009sh}. 
This expresses the full Pomeron Green function $G\Lb\nu,y\Rb$ in $\nu,y$-representation, 
as the sum over the class of diagrams in
\fig{fren} as;

\bea
G\Lb\nu,y\Rb=g\Lb\nu\Rb-g\Lb\nu\Rb\int^y_0\!\!\! dy_1\!\int^{y_1}_0\!\!\! dy_2 \,m\Lb\nu,y_{12}\Rb G\Lb\nu,y_2\Rb\label{SDequationclear}\eea

where $m\Lb\nu,y_{12}\Rb$ is the Pomeron self mass ($y_{12}=y_1-y_2$), which sits between the rapidity values $y_1$ for the upper rapidity limit,
 and $y_2$ for the lower rapidity limit (see \fig{fren} (a)),
 and is
given by the expression;

\bea
m\Lb\nu,y_{12}\Rb&&=\f{1}{16}
\intf\!\!\! d\nu_1h\Lb\nu_1\Rb G\Lb\nu_1,y_{12}\Rb\!\intf\!\!\! d\nu_2h\Lb\nu_2\Rb G\Lb\nu_2,y_{12}\Rb|\Ga\Lb\nu,\nu_1,\nu_2\Rb|^2
\,e^{\Lb \omega\Lb\nu_1\Rb+\omega\Lb\nu_2\Rb-\omega\Lb\nu\Rb\Rb y_{12}}\hspace{1.5cm}\label{selfmassclear}
\eea

The pre-factor of $1/16$ in \eq{selfmassclear} divides by the order of the symmetry group, so that identical diagrams
 are only counted once (see ref. \cite{Miller:2009ca} for a full explanation).
$\Ga\Lb\nu,\nu_1,\nu_2\Rb$ is the triple Pomeron vertex for the splitting of the Pomeron with the conformal variable $\nu$ into two daughter Pomerons
with conformal variables $\nu_1$ and $\nu_2$ which form the branches of the loop. The splitting vertex is the complex conjugate of the merging vertex,
 so the squared absolute value of the vertex in \eq{selfmassclear} is the product of the splitting and the re-merging vertex, forming the loop.\\
 
The Schwinger Dyson \eq{SDequationclear} forms an iterative sum,
which after expanding  takes the following form;

\begin{subequations}\label{SDequation}
\bea
G\Lb\nu,y\Rb
&&=g\Lb\nu\Rb\sum^\infty_{n=0}\Lb -1\Rb^ng^{n}\Lb \nu\Rb\prod^n_{i=1}\int^{y_2^{i+1}}_0\!\!\! dy_1^i\!\int^{y_1^i}_0\!\!\! dy_2^i \,m\Lb\nu,y_{12}^i\Rb;\hspace{1.5cm}
y_2^{n+1}=y;\label{SchwingerDyson}\\
m\Lb\nu,y_{12}^i\Rb&&=\f{1}{16}
\intf\!\!\! d\nu_1h\Lb\nu_1\Rb G\Lb\nu_1,y_{12}^i\Rb\!\intf\!\!\! d\nu_2h\Lb\nu_2\Rb G\Lb\nu_2,y_{12}^i\Rb|\Ga\Lb\nu,\nu_1,\nu_2\Rb|^2
\,e^{\Lb \omega\Lb\nu_1\Rb+\omega\Lb\nu_2\Rb-\omega\Lb\nu\Rb\Rb y_{12}^i}\hspace{1.5cm}\label{selfmass}
\eea
\end{subequations}

where $m\Lb\nu,y_{12}^i\Rb$ is the Pomeron self mass which sits between the rapidity values $y_1^i$ and $y_2^i$ in \fig{fren} (b), and
  $y^i_{12}=y^i_1-y^i_2$. The  $G\Lb\nu_1,y_{12}^i\Rb$ and $G\Lb\nu_2,y_{12}^i\Rb$  in \eq{selfmass} are the same Pomeron propagators given by
the infinite Schwinger-Dyson sum of \eq{SchwingerDyson}. From this it becomes clear how plugging \eq{selfmass} into \eq{SchwingerDyson} generates the non-closed sum over
the never ending spectrum of Pomeron loop diagrams. The integration
limits are due to the upper rapidity value $y_1^i$ in \fig{fren} which cannot exceed the  lower rapidity value $y_2^{i+1}$ of the next self mass interaction  above it.
The strategy for deriving a closed expression which is a solution to \eq{SchwingerDyson} is the following.\\

First consider  the sum over diagrams of the type shown in \fig{fren} (a), which contains $n$ Pomeron loops in series. The sum over all such diagrams from $n=0$ to infinity is described by
 the Schwinger Dyson \eq{SchwingerDyson} by replacing $m\Lb\nu,y_{12}^i\Rb$ with the Pomeron loop $m_0\Lb\nu,y_{12}^i\Rb$
which yields;

\begin{subequations}\lab{SchwingerDysonzero}
\bea
G_0\Lb\nu,y|\mbox{\fig{fren} (a)}\Rb
&&=g\Lb\nu\Rb\sum^\infty_{n=0}\Lb -1\Rb^ng^{n}\Lb \nu\Rb\prod^n_{i=1}\int^{y_2^{i+1}}_0\!\!\! dy_1^i\!\int^{y_1^i}_0\!\!\! dy_2^i \,m_0\Lb\nu,y_{12}^i\Rb\label{SchwingerDyson0}\\
m_0\Lb\nu,y_{12}^i\Rb&&=\f{1}{16}
\intf\!\!\! d\nu_1h\Lb\nu_1\Rb g\Lb\nu_1\Rb\!\intf\!\!\! d\nu_2h\Lb\nu_2\Rb g\Lb\nu_2\Rb|\Ga\Lb\nu,\nu_1,\nu_2\Rb|^2
\,e^{\Lb \omega\Lb\nu_1\Rb+\omega\Lb\nu_2\Rb-\omega\Lb\nu\Rb\Rb y_{12}^i}\hspace{1cm}\label{selfmass0}\eea
\end{subequations}

where  $m_0\Lb\nu,y_{12}^i\Rb$ is found from \eq{selfmass}  by replacing the renormalized propagators $G\Lb\nu_1,y_{12}^i\Rb$ and $G\Lb\nu_2,y_{12}^i\Rb$,
 with the bare propagators $g\Lb\nu_1\Rb$ and $g\Lb\nu_2\Rb$.
From the Korchemsky expression for the triple Pomeron vertex  \cite{Korchemsky:1997fy}, it was found in ref. \cite{Miller:2009ca,Miller:2009} that there are two
main contributions to the triple Pomeron vertex which are given by the following asymptotes;

\begin{subequations}\label{tpv}
\bea
&&|\Ga\Lb i\nu=\h,i\nu_1=0,i\nu_2=0\Rb|^2\!=\!\f{\Lb 4\pi\Rb^6\bas^4}{N_c^4}\f{1}{1/4+\nu^2}\label{tpv2}\\
&&|\Ga\Lb\nu,i\nu_1=\h,i\nu_2=\h\Rb|^2\!=\!\h\Lb 4\pi\Rb^6\bas^4\Lb 1-\f{1}{N_c^2}\Rb^2\!\!\chi\Lb\nu\Rb\!\Lb \f{1}{4}+\nu^2\Rb^3\!\!\f{ 1-i\nu_1-i\nu_2}{\Lb 1/2-i\nu_1\Rb^2\Lb 1/2-i\nu_2\Rb^2}\hspace{1.6cm}\label{tpv1}\\
&&\chi\Lb\nu\Rb\!=\!\Re e\Lb\psi\Lb 1\Rb-\psi\Lb\h+i\nu\Rb\Rb
\eea
\end{subequations}

\eq{tpv2} leads to the contribution to the Pomeron loop amplitude of \eq{selfmass0} which is equivalent to $2$ non interacting Pomerons, with renormalized Pomeron vertices.
 This particular contribution to  $m_0\Lb\nu,y_{12}^i\Rb$
  gives a vanishing result for $n>1$ in \eq{SchwingerDyson0}. This makes sense since \eq{SchwingerDyson0} describes the sum over $n$ Pomeron loops in series shown in \fig{fren} (a).
Therefore
the phenomena where the branches of the loop span the entire rapidity gap between the projectile and target to become 2 independent partons, cannot occur with more than 1 loop in series.
With this in mind the contribution to the vertex of \eq{tpv2} is postponed until the non interacting
Pomeron solution is discussed later on in \sec{sNIP}. For now, inserting the asymptote of \eq{tpv1} into \eq{selfmass0}, and then introducing the definitions given in \eq{hla} (where $g\Lb\nu\Rb$ can be cast as $1/16\Lb 1/2+i\nu\Rb^2\Lb 1/2-i\nu\Rb^2\,$ );

\bea
m_0\Lb\nu,y_{12}^i\Rb&&=\f{\pi^6}{2}\!\!\Lb 1-\f{1}{N_c^2}\Rb^2\!\!\Lb\f{1}{4}+\nu^2\Rb^3\!\!\chi\Lb\nu\Rb\label{selfmass0a}\\
&&\times\intf\!\! \f{d\nu_1\,\nu_1^2}{\Lb \h+i\nu_1\Rb^2\Lb \h-i\nu_1\Rb^4}\!\intf\!\!
\f{d\nu_2\,\nu_2^2\Lb 1-i\nu_1-i\nu_2\Rb}{\Lb \h+i\nu_2\Rb^2\Lb \h-i\nu_2\Rb^4}
\,e^{\Lb \omega\Lb\nu_1\Rb+\omega\Lb\nu_2\Rb-\omega\Lb\nu\Rb\Rb y_{12}^i}\nn\eea

The $\nu_1,\nu_2$ integrals are solved
 by closing the contour over the upper half plane and summing over the residues at $i\nu_1,i\nu_2=1/2$, taking into account the poles which stem from
 $\omega\Lb\nu_{1,2}\Rb\xrightarrow{i\nu_{1,2}\to 1/2}-\bas/\Lb 1/2-i\nu_{1,2}\Rb$ such that \eq{selfmass0a} becomes

\bea
m_0\Lb\nu,y_{12}^i\Rb&&=\f{\pi^6}{2}\!\Lb 1-\f{1}{N_c^2}\Rb^2\!\!\Lb\f{1}{4}+\nu^2\Rb^3\!\!\chi\Lb\nu\Rb
\oint_C\f{d\nu_1\,\nu_1^2}{\Lb \h+i\nu_1\Rb^2\!\Lb \h-i\nu_1\Rb^4}\oint_C\f{d\nu_2\,
\nu_2^2\Lb 1-i\nu_1-i\nu_2\Rb}{\Lb \h+i\nu_2\Rb^2\!\Lb \h-i\nu_2\Rb^4}\nn\\
&&\times e^{\Lb -\bas\Lb 1/2-i\nu_1\Rb^{-1}-\bas\Lb 1/2-i\nu_2\Rb^{-1}-\omega\Lb\nu\Rb\Rb y_{12}^i}\nn\\
\nn\\
&&=\f{\pi^6}{2}\Lb 1-\f{1}{N_c^2}\Rb^2\!\!\Lb\f{1}{4}+\nu^2\Rb^3\!\!\chi\Lb\nu\Rb
\oint_C\f{d\nu_1\,\nu_1^2}{\Lb \h+i\nu_1\Rb^2\Lb \h-i\nu_1\Rb^2}\Lb\f{-1}{\bas}\f{d}{dy_{12}^i}\Rb^2
e^{ -\bas\Lb 1/2-i\nu_1\Rb^{-1}y_{12}^i}\nn\\
&&\times\oint_C\f{d\nu_2\,
\nu_2^2\Lb 1-i\nu_1-i\nu_2\Rb}{\Lb \h+i\nu_2\Rb^2\Lb \h-i\nu_2\Rb^4}
\,e^{\Lb -\bas\Lb 1/2-i\nu_2\Rb^{-1}-\omega\Lb\nu\Rb\Rb y_{12}^i}\label{selfmass0b}
\eea

It is instructive to change the $\nu_1$ integration variable to $u=\bas/\Lb 1/2-i\nu_1\Rb+\bas/\Lb 1/2-i\nu_2\Rb$, such that the jacobian cancels the remaining singularity
which stems from $\Lb 1/2-i\nu_1\Rb^{-2}$ in \eq{selfmass0b}. Integrating over $u$ yields the derivative of the Dirac delta function $2\pi i\de^{(2)}\Lb y_1^i-y_2^i\Rb$.
After taking the residue at $i\nu_1=1/2$ the $\nu_2$ integral reduces to $\oint_Cd\nu_2\,
\nu_2^2/\Lb \h+i\nu_2\Rb^2\Lb \h-i\nu_2\Rb^3=2\pi i/4$. Over all, after re-arranging the order of derivatives so that there are no derivatives of the delta
function in the integrand, one finds for the Pomeron loop amplitude the following expression;

\bea
m_0\Lb\nu,y_{12}^i\Rb&&=a\Lb\f{1}{4}+\nu^2\Rb^3\chi\Lb\nu\Rb\omega^2\Lb\nu\Rb e^{-\omega\Lb\nu\Rb y_{12}^i}\de\Lb y_1^i-y_2^i\Rb;\hspace{1cm}a=\bas\Lb 1-\f{1}{N_c^2}\Rb^2\label{selfmass01}
\eea

Finally inserting \eq{selfmass01} into \eq{SchwingerDyson0},  the integrations over the rapidity variables are trivially solved thanks to the Dirac delta function.
Thus one derives the following expression for the sum over the class of diagrams of \fig{fren} (a) with $n$ consecutive loops ;

\bea
G_0\Lb\nu,y|\mbox{\fig{fren} (a)}\Rb
&&=g\Lb\nu\Rb\sum^\infty_{n=0}\f{\Lb -1\Rb^n}{n!}
\Lb g\Lb \nu\Rb\,a\,\Lb 1/4+\nu^2\Rb^3\chi\Lb\nu\Rb\omega^2\Lb\nu\Rb\,y\,\Rb^n\nn\\
&&=g\Lb\nu\Rb\exp\Lb-\f{a}{16}
\Lb \f{1}{4}+\nu^2\Rb\chi\Lb\nu\Rb\omega^2\Lb\nu\Rb y\Rb\label{SchwingerDyson01}\eea

Substituting for the bare propagator $g\Lb\nu\Rb$ that appears in \eq{A0}, the renormalized one derived in \eq{SchwingerDyson01},
leads to the scattering amplitude which is equivalent to the replacement;

\bea
\omega\Lb\nu\Rb\to\widetilde{\omega}\Lb\nu\Rb=\omega\Lb\nu\Rb+\zeta\Lb\nu\Rb;\hspace{1cm}\zeta\Lb\nu\Rb=
-\f{a}{16}\Lb\f{1}{4}+\nu^2\Rb\chi\Lb\nu\Rb\omega^2\Lb\nu\Rb
\label{replacement}\eea

Therefore the sum over the class of diagrams in \fig{fren} (a)  leads
directly to the renormalized Pomeron intercept given by \eq{replacement}.
Phrased differently, the sum over the class of diagrams of \fig{fren} (a) is found by replacing $\omega\Lb\nu\Rb$
in  \eq{A0} with \eq{replacement}.
The bold lines in \fig{fren} (b), represent the sum over the class of diagrams in \fig{fren} (a). As such \fig{fren} (b) is a string of loops, where the bold lines
which form the loops are themselves a string of loops. The bold lines are reggeons, with the renormalized intercept $\tom\Lb\nu\Rb$
derived in \eq{replacement}. With this in mind,
the sum over the class of diagrams of \fig{fren} (b) is derived from \eq{SchwingerDysonzero}, by replacing the intercepts $\omega\Lb \nu_1\Rb$ and $\omega\Lb\nu_2\Rb$  
with $\tom\Lb \nu_1\Rb$ and $\tom\Lb\nu_2\Rb$  found in \eq{replacement}, which yields;

\begin{subequations}\label{SD2}
\bea
G\Lb\nu,y|\mbox{\fig{fren} (b)}\Rb
&&=g\Lb\nu\Rb\sum^\infty_{n=0}\Lb -1\Rb^ng^{n}\Lb \nu\Rb\prod^n_{i=1}\int^{y_2^{i+1}}_0\!\!\! dy_1^i\!\int^{y_1^i}_0\!\!\! dy_2^i \,m\Lb\nu,y_{12}^i\Rb;\label{SchwingerDyson2}\\
\nn\\
m\Lb\nu,y_{12}^i\Rb&&=\f{1}{16}
\intf\!\!\! d\nu_1h\Lb\nu_1\Rb g\Lb\nu_1\Rb\!\intf\!\!\! d\nu_2h\Lb\nu_2\Rb g\Lb\nu_2\Rb|\Ga\Lb\nu,\nu_1,\nu_2\Rb|^2
\,e^{\Lb \widetilde{\omega}\Lb\nu_1\Rb+\widetilde{\omega}\Lb\nu_2\Rb-\omega\Lb\nu\Rb\Rb y_{12}^i}\hspace{1.5cm}\label{selfmass2}\eea
\end{subequations}

Following the same arguments used above, the integrations in \eq{selfmass2} are solved by closing the $\nu_1,\nu_2$ contours
over the upper half plane and summing over the residues at $i\nu_1,i\nu_2=1/2$. After introducing the definitions given in \eq{hla} and taking into account the
singularities that stem from the triple Pomeron vertex of \eq{tpv1}, and the asymptotes;

\bea
 \omega\Lb\nu\Rb
\xrightarrow{i\nu\to 1/2}\f{-\bas}{ \h-i\nu}; \hspace{0.5cm}\zeta\Lb\nu\Rb\xrightarrow{i\nu\to 1/2}\f{a}{16}\f{\bas^2}{\Lb \h-i\nu\Rb^2}\lab{asymptotes}\eea

\eq{selfmass2} becomes;

\bea
m\Lb\nu,y_{12}^i\Rb&&=\f{\pi^6}{2}\!\Lb 1-\f{1}{N_c^2}\Rb^2\!\!\Lb\f{1}{4}+\nu^2\Rb^3\!\!\chi\Lb\nu\Rb\nn\\
&&\times\oint_C\f{ d\nu_1\,\nu_1^2}{\Lb \h+i\nu_1\Rb^2\Lb \h-i\nu_1\Rb^4}
\exp\Lb \f{-\bas}{ \h-i\nu_1}y_{12}^i+\f{a}{16}\f{\bas^2}{\Lb \h-i\nu_1\Rb^2} y_{12}^i\Rb\nn\\
&&\times\oint_C\f{ d\nu_2\,\nu_2^2\Lb 1-i\nu_1-i\nu_2\Rb}{\Lb \h+i\nu_2\Rb^2\Lb \h-i\nu_2\Rb^4}
\,e^{\Lb\widetilde{\omega}\Lb\nu_2\Rb-\omega\Lb\nu\Rb\Rb y_{12}^i}\nn\\
\nn\\
\nn\\
&&=\f{\pi^6}{2}\!\Lb 1-\f{1}{N_c^2}\Rb^2\!\!\Lb\f{1}{4}+\nu^2\Rb^3\!\!\!\chi\Lb\nu\Rb\!\nn\\&&\times
\oint_C\!\f{ d\nu_1\,\nu_1^2}{\Lb \h+i\nu_1\Rb^2\Lb \h-i\nu_1\Rb^2}\Lb\f{-1}{\bas}\f{d}{dy_{12}^i}\Rb^2\!\exp\Lb   \f{a}{16} y_{12}^i\f{d^2}{d^2y_{12}^i}\Rb\!\exp\Lb \f{-\bas}{ \h-i\nu_1}y_{12}^i\Rb\nn\\
&&\times\oint_C\f{ d\nu_2\,\nu_2^2\Lb 1-i\nu_1-i\nu_2\Rb}{\Lb \h+i\nu_2\Rb^2\Lb \h-i\nu_2\Rb^4}
\,e^{\Lb\widetilde{\omega}\Lb\nu_2\Rb-\omega\Lb\nu\Rb\Rb y_{12}^i}\hspace{1.5cm}\label{selfmass3}\eea

Changing the integration variable to $w=\bas/\Lb 1/2-i\nu_1\Rb+\tom\Lb\nu_2\Rb$ and integrating over $w$ yields the Dirac delta function $2\pi i\de\Lb y_1^i-y_2^i\Rb$,
acted on by the derivatives with respect to $y_{12}^i$ which appear in the integrand. After taking the residue at $i\nu_1=1/2$ the $\nu_2$ integral reduces to
$\oint_Cd\nu_2\,\nu_2^2/\Lb \h+i\nu_2\Rb^2\Lb \h-i\nu_2\Rb^3=2\pi i/4$. After rearranging the order of derivatives so that there is no derivative of the Dirac Delta function
in the final expression;

\begin{subequations}\lab{selfmass4}
\bea
m\Lb\nu,y_{12}^i\Rb&&=a\Lb\f{1}{4}+\nu^2\Rb^3\chi\Lb\nu\Rb\omega^2\Lb\nu\Rb\,e^{-\omega\Lb\nu\Rb y_{12}^i}\left\{\exp\Lb-\f{a}{16}\omega^2\Lb\nu\Rb y_{12}^i\Rb+\phi\Lb\nu\Rb\right\}
\de\Lb y_1^i-y_2^i\Rb\lab{selfmass4a}\\
\phi\Lb\nu\Rb&&=\f{1-\sqrt{1-4\varphi\Lb\nu\Rb}-2\varphi\Lb\nu\Rb}{2\sqrt{1-4\varphi\Lb\nu\Rb}\varphi^2\Lb\nu\Rb};\hspace{0.5cm}\varphi\Lb\nu\Rb=\f{a}{16}\omega\Lb\nu\Rb\lab{selfmass4b}
\eea
\end{subequations}

Finally inserting \eq{selfmass4a} into \eq{SchwingerDyson2} and evaluating all the integrations over the rapidity variables leads to the
following result for the sum over the class of diagrams  shown  in \fig{fren} (b);

\bea
G\Lb\nu,y|\mbox{\fig{fren} (b)}\Rb =g\Lb\nu\Rb \exp\left\{ -\f{a}{16}\Lb\f{1}{4}+\nu^2\Rb\chi\Lb\nu\Rb\omega^2\Lb\nu\Rb\Lb 1+\phi\Lb\nu\Rb\Rb\,y\,\right\}\label{SchwingerDyson3}\eea

Using exactly the same arguments which led to \eq{replacement}, the sum over the class of diagrams with $n$ consecutive self mass terms  shown in \fig{fren} (b), is achieved by replacing
 $\omega\Lb\nu\Rb$ that appears in \eq{A0}, with the following renormalized Pomeron intercept;

\bea
\omega\Lb\nu\Rb\to\widetilde{\tom}\Lb\nu\Rb=\omega\Lb\nu\Rb+\zeta\Lb\nu\Rb\Lb 1+\phi\Lb\nu\Rb\Rb\lab{renormalizedPomeronintercept}\eea

If the same treatment is repeated and another deeper level of loops are introduced, the Pomeron intercepts $ \omega\Lb\nu_1\Rb$ and $\omega\Lb\nu_2\Rb$ in \eq{selfmass0} are replaced by \eq{renormalizedPomeronintercept}.
The $\nu_1,\nu_2$ integrals are solved in the same way by closing the contour over the upper half plane and summing over the residues at $i\nu_1,i\nu_2=1/2$.
Fortunately, $\phi\Lb\nu_{1,2}\Rb$ vanishes as $i\nu_{1,2}\to 1/2$, so it gives no contribution to the sum over residues. This means that the steps from \eq{SD2} to \eq{SchwingerDyson3}
are identical, and the same expression of \eq{renormalizedPomeronintercept} for the renormalized Pomeron intercept is derived. Therefore every time
a deeper level of loops is introduced, one always arrives at \eq{renormalizedPomeronintercept} for the renormalization of the Pomeron intercept. Therefore the following
expression derived in \eq{SchwingerDyson3}

\bea
G\Lb\nu,y\Rb =g\Lb\nu\Rb \exp\left\{ -\f{a}{16}\Lb\f{1}{4}+\nu^2\Rb\chi\Lb\nu\Rb\omega^2\Lb\nu\Rb\Lb 1+\phi\Lb\nu\Rb\Rb\,y\,\right\}\label{SchwingerDysonsolution}\eea

is the solution to the Schwinger Dyson equation \eq{SchwingerDyson} which describes the sum over the class of Pomeron loop diagrams in \fig{fren}. The solution is equivalent to
the replacement of the Pomeron intercept $\omega\Lb\nu\Rb$, with the renormalized  Pomeron intercept $\widetilde{\tom}\Lb\nu\Rb$ derived in \eq{renormalizedPomeronintercept}.

\section{The non interacting Pomeron solution}
\label{sNIP}

\DOUBLEFIGURE[h]{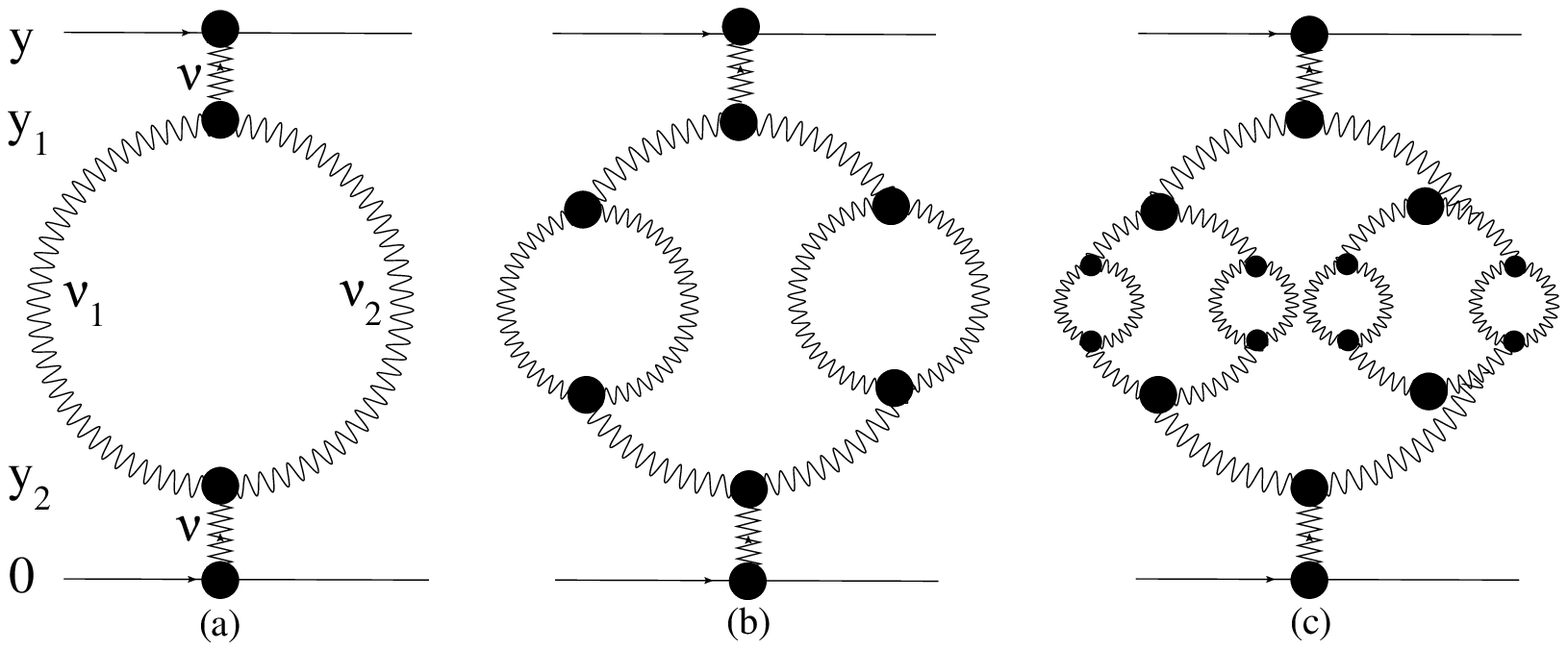,width=85mm,height=40mm}{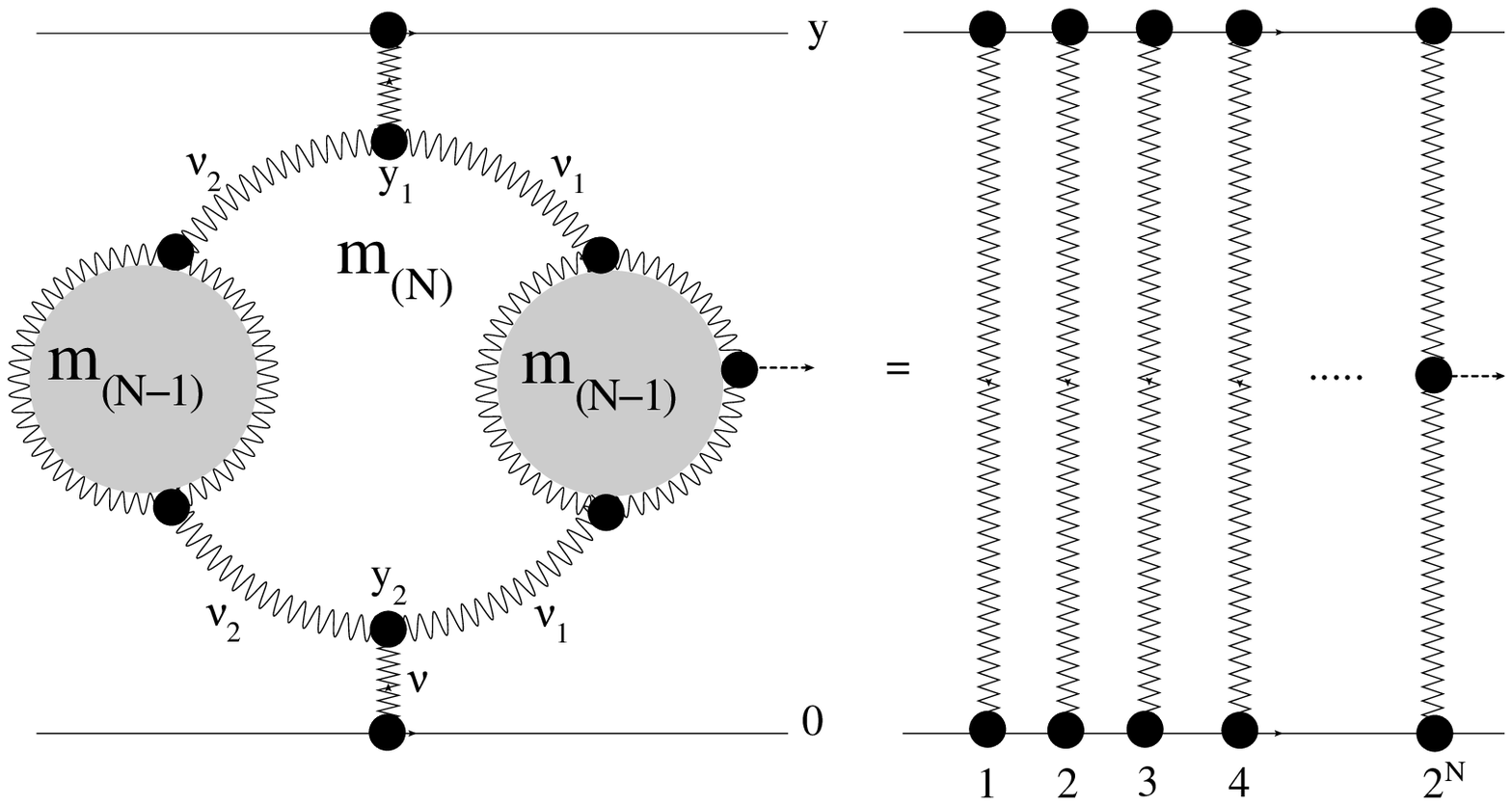,width=85mm,height=40mm}
{ The special class of symmetric Pomeron loop diagrams taken into account in the sum over Pomeron loops. (a) is the diagram with $N=1$ generation of loops, (b) has $N=2$ generations of loops and (c)
has $N=3$ generations of loops.
\label{fsopl}}{The $N$ generation diagram which stems from the simple loop giving
birth to two sets of $N-1$ generations of loops, is equivalent to the diagram of $2^N$ non interacting Pomerons with renormalized Pomeron vertices.\label{fpictN1} }

The next part of the discussions is focussed on the non interacting Pomeron solution, which stems from diagrams of the type shown in \fig{fsopl}. Loops which become non interacting Pomerons do not contribute to consecutive loops in
\fig{fren} (a), because
the phenomena where the branches of the loop stretch and span the whole rapidity gap between the projectile and target, can't happen for more than one loop in series.
Therefore this solution cannot be treated in the context of the Schwinger Dyson \eq{SchwingerDyson} for $n > 1$, and requires a separate approach.
Using the same conventions, the scattering amplitude of \fig{fsopl} (a) is given by the expression;

 \begin{subequations}\label{loop}
 \bea
&&A_{(1)}\Lb y |\mbox{\fig{fsopl} (a)}\Rb=\f{\as^2}{4}\!\If\!\!\!d\nu h\Lb\nu\Rb g^2\Lb\nu\Rb e^{\omega\Lb\nu\Rb y}m_{(1)}\Lb\nu,y\Rb E_\nu E^\prime_{-\nu}\label{loop1}\\
&&m_{(1)}\Lb\nu,y\Rb=\f{1}{16}\If\!\!\! d\nu_1 h\Lb\nu_1\Rb g\Lb\nu_1\Rb\If\!\!\! d\nu_2 h\Lb\nu_2\Rb g\Lb\nu_2\Rb|\Ga\Lb\nu|\nu_1,\nu_2\Rb|^2\label{loop2}\\
&&\times\int^y_0 dy_1\!\int^{y_1}_0 dy_2 e^{\Lb\omega\Lb\nu_1\Rb+\omega\Lb\nu_2\Rb-\omega\Lb\nu\Rb\Rb y_{12}}\nn
 \eea
 \end{subequations}

 where $y_{12}=y_1-y_2$ is the rapidity gap which the loop fills (see \fig{fsopl} (a)). Now inserting the asymptote of \eq{tpv2} for the triple Pomeron vertex for the region where $\nu_1,\nu_2$ are close to zero,
 one can substitute for the BFKL eigenfunctions $\omega\Lb\nu_1\Rb$ and $\omega\Lb\nu_2\Rb$ the expansion of \eq{BFKLexpansion} and integrate over $\nu_1$ and $\nu_2$ using the
 method of steepest descents, which gives the expression \cite{Miller:2009ca,Miller:2009};

\bea
m_{(1)}\Lb\nu,y\Rb&&=\f{b }{\Lb 1/4+\nu^2\Rb }\int^y_0 dy_1\!\int^{y_1}_0 dy_2 \f{e^{\Lb 2\omega\Lb 0\Rb-\omega\Lb\nu\Rb\Rb y_{12}}}{y_{12}^3};\hspace{1cm}
b=\f{2^{10}\bas^4}{N_c^4\pi\Lb\omega^{\prime\prime}\Lb 0\Rb\Rb^3}\,.\label{fourth}
\eea

Inserting \eq{fourth} back into \eq{loop1}, the $\nu$ integration can be solved by closing the contour over the upper half plane and summing over the residues at $i\nu=1/2$, taking into account the
singularity which stems from $\omega\Lb\nu\Rb$ given in \eq{asymptotes}. This leads to the following contribution to the scattering amplitude  \cite{Miller:2009ca,Miller:2009};

\bea
A_{(1)}\Lb y|\mbox{\fig{fsopl} (a)}\Rb&&=\f{\bas b}{2^9\pi N_c^2}y\Lb\f{-1}{\bas}\f{d}{dy}\Rb^3\f{e^{2\omega\Lb 0\Rb y}}{y^3}\label{A2}
\eea

\eq{A2} is equivalent to 2 non interacting Pomerons, with renormalized Pomeron vertices.
As explained above, \fig{fsopl} (b) stems from \fig{fsopl} (a) when each branch of the loop gives birth to a secondary loop leading to the two ``second generation" of loops in  \fig{fsopl} (b).
 In the same way when the second generation loops in \fig{fsopl} (b) each give birth to two loops, this leads to 4 ``third generation" of loops in \fig{fsopl} (c). Continuing with this evolution, the entire spectrum of symmetric $N$ generation diagrams can be generated for all $N$ . The scattering amplitude with $N$ generations of loops shown in \fig{fpictN1}
is the generalization of \eq{loop1}, namely  \cite{Miller:2009ca,Miller:2009};

\bea
A_{(N)}\Lb y|\mbox{\fig{fpictN1}}\Rb&&=\f{\as^2}{4}\If\!\!\! d\nu h\Lb\nu\Rb g^2\Lb\nu\Rb e^{\omega\Lb\nu\Rb y}m_{(N)}\Lb\nu,y\Rb E_\nu E^\prime_{-\nu} \label{AN}\eea

where $m_{(N)}\Lb\nu,y\Rb$ is the contribution of the $N$ generations of loops in \fig{fpictN1}. In refs.  \cite{Miller:2009ca,Miller:2009} a detailed explanation of how to calculate $m_{(N)}\Lb\nu,y\Rb$ was given. This
is based on the observation that \fig{fpictN1} is equivalent to the simple loop diagram of \fig{fsopl} (a) when each branch in the loop gives birth to a set of $N-1$ generations of loops. This means
that to write the expression for $m_{(N)}\Lb\nu,y\Rb$, all that is needed is to modify the propagators for the branches of the loop in the expression of \eq{loop2} as;

\bea
g\Lb\nu_1\Rb\to g\Lb\nu_1\Rb m_{(N-1)}\Lb\nu_1,y_{12}\Rb g\Lb\nu_1\Rb;\hspace{0.5cm}
g\Lb\nu_2\Rb\to g\Lb\nu_2\Rb m_{(N-1)}\Lb\nu_2,y_{12}\Rb g\Lb\nu_2\Rb\label{lamodified}\eea

After implementing \eq{lamodified} in \eq{loop2}, one arrives at the following  amplitude for the set of $N$ generations of loops;

\bea
m_{(N)}\Lb\nu|y,\de y_H\Rb&&=\f{1}{16}\If\!\!\! d\nu_1 h\Lb\nu_1\Rb g\Lb\nu_1\Rb\If\!\!\! d\nu_2 h\Lb\nu_2\Rb g\Lb\nu_2\Rb|\Ga\Lb\nu|\nu_1,\nu_2\Rb|^2\label{loopN}\\
&&\times\int^y_0dy_1\!\int^{y_1}_0dy_2 e^{\Lb\omega\Lb\nu_1\Rb+\omega\Lb\nu_2\Rb-\omega\Lb\nu\Rb\Rb y_{12}}
m_{(N-1)}\Lb\nu_1,y_{12}\Rb m_{(N-1)}\Lb\nu_2,y_{12}\Rb\nn
 \eea

\eq{loopN} forms an iterative expression. Using the technique of proof by induction, the following formula  derived in refs.  \cite{Miller:2009ca,Miller:2009} for \eq{loopN} can be proved;

\begin{subequations}\label{dN1}
\bea
m_{(N)}\Lb\nu,y\Rb&&=
\f{\Lb b b^{\,\prime}\Rb^{2^{[N-1]}}}{b^{\,\prime}}\Lb \f{1}{4}+\nu^2\Rb^3\chi\Lb\nu\Rb \int^y_0 dy_1\int^{y_1}_0dy_2\, y_{12}^{2^{[N-1]}-1}
\left\{\!\Lb\f{-1}{\bas}\f{d}{dy_{12}}\Rb^3\!\f{e^{2\omega\Lb 0\Rb y_{12}}}{y_{12}^3}\!\right\}^{2^{[N-1]}}\label{dN3}\\
\nn\\
b&&=\f{2^{10}\bas^4}{N_c^4\pi[\omega^{\,\prime\prime}\Lb 0\Rb]^3};\hspace{0.5cm}
b^\prime=\f{\bas^2}{2^{11}}\Lb 1-\f{1}{N_c^2}\Rb^2.\label{fullsetofconstants}
\eea
\end{subequations}

Finally after inserting \eq{dN1} into \eq{AN}, one finds the following scattering amplitude for the diagram of \fig{fpictN1} with $N$ generations of loops;

\bea
A_{(N)}\Lb y|\mbox{\fig{fpictN1}}\Rb&&=\f{\bas\,}{2^9 N_c^2\pi}
\f{\Lb b b^\prime\Rb^{2^{[N-1]}}}{b^\prime}y^{2^{[N-1]}}
\left\{\!\!\Lb\f{-1}{\bas}\f{d}{dy}\Rb^3\!\!\f{e^{2\omega\Lb 0\Rb y}}{y^3}\!\!\right\}^{2^{[N-1]}}\lab{AN3}\eea

\eq{AN3}  is equivalent to $2^N$ non interacting Pomerons, with renormalized Pomeron vertices
shown pictorially in \fig{fpictN1}. This follows from the observation that \eq{AN3} can be recast in the form;

\bea
A_{(N)}\Lb y|\mbox{\fig{fpictN1}}\Rb\equiv \ka_{\mbox{\tiny{\it (N)}}}\,e^{2^N\omega\Lb 0\Rb y}\label{noninteratingpomerons}\eea

where the coefficient $\ka_{\mbox{\tiny{\it (N)}}}$
 contains the set of renormalized Pomeron vertices. \eq{AN3} describes the scattering amplitude with
$2^{N-1}$ loops, which is equivalent to $2^N$ non interacting Pomerons. This formula can be generalized
 to the scattering amplitude of the symmetric diagram which contains $n$ loops, or equivalently $2n$ non interacting Pomerons, namely;

\bea
A_{(n)}\Lb y\Rb&&=\f{\bas\,}{2^9 N_c^2\pi}
\f{\Lb b b^{\,\prime}\Rb^n}{b^{\,\prime}}y^n
\left\{\!\!\Lb\f{-1}{\bas}\f{d}{dy}\Rb^3\!\!\f{e^{2\omega\Lb 0\Rb y}}{y^3}\!\!\right\}^n\lab{AN3generalized}\eea

The sum over the complete set of symmetric Pomeron loop diagrams, is achieved by evaluating the sum
$\sum^\infty_{n=0} A_{(n)}\Lb y\,|\,\mbox{\eq{AN3generalized}}\Rb$.
In the outcome formula, the intercept $\omega\Lb 0\Rb$ should be replaced with $\widetilde{\tom}\Lb 0\Rb$ of \eq{renormalizedPomeronintercept}.
This leads to the sum over diagrams with independent Pomeron exchanges, where the Pomerons are replaced by the superposition of loop series shown in \fig{fren}, described by the
Schwinger Dyson equation. In this formalism the full sum over symmetric  Pomeron loop diagrams is given by;

\bea
\sum^\infty_{n=1}A_{(n)}\Lb y\Rb&&=\f{\bas\,}{2^9 N_c^2\pi} y b \Lb\f{-1}{\bas}\f{d}{dy}\Rb^3\!\!\f{e^{2\widetilde{\tom}\Lb 0\Rb y}}{y^3}\label{sumoverPomeronloops}\\
&&\overline{\hspace{0.4cm}1+b b^{\,\prime}y\, \Lb\f{-1}{\bas}\f{d}{dy}\Rb^3\!\!\f{e^{2\widetilde{\tom}\Lb 0\Rb y}}{y^3}\hspace{0.4cm}}\nn
\eea

\vspace{0.1cm}

\section{Results}
\label{sR}

\FIGURE[h]{\begin{minipage}{70mm}
\centerline{\epsfig{file=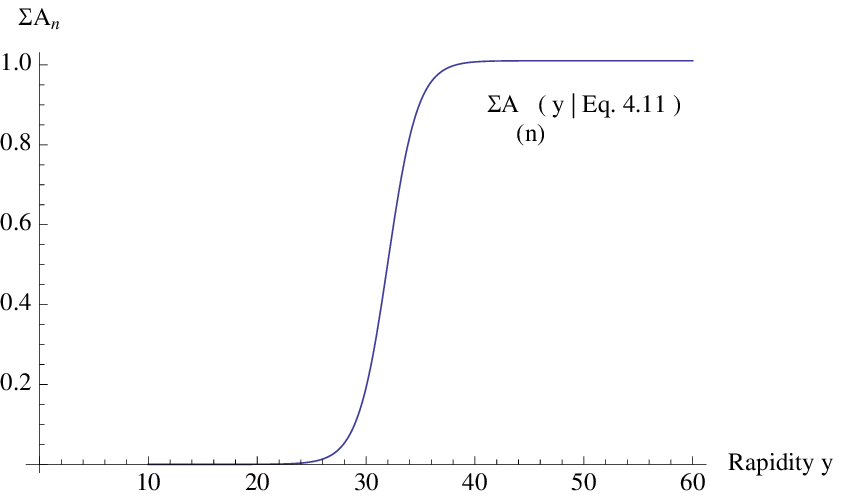,width=70mm}}
\end{minipage}
\caption{ The energy dependence of the sum over symmetric Pomeron loop diagrams $\sum^\infty_{n=1}A_{(n)}\Lb y|\mbox{\eq{sumoverPomeronloops}}\Rb$  } \label{fsed2} }

The energy dependence of the sum over Pomeron loop diagrams
 is shown in \fig{fsed2}. The curve approaches the black disk limit, so that
 unitarity is preserved. If the bare Pomeron intercept $\omega\Lb 0\Rb$ is used in \eq{sumoverPomeronloops}
instead of the renormalized one, the graph of \fig{fsed2} is unaffected. This indicates that the dominant contribution
to the sum over Pomeron loop diagrams comes from the diagrams which are equivalent to non interacting Pomerons, with renormalized Pomeron vertices.
The class of diagrams of \fig{fren} leading to the renormalized Pomeron intercept
give a negligible contribution in comparison.\\

The formula of \eq{sumoverPomeronloops} requires explanation. It is tempting to think that
\eq{sumoverPomeronloops} is the scattering amplitude, however a closer look reveals that this formula is still just the
sum over a special class of loop diagrams, as will now be explained.

 \section{Conclusions and discussion}
 \label{cad}

The following discussion is a summary of the formalism adopted in this paper, for the summation of Pomeron loops.
There are two distinct ways of summing over Pomeron loops, and both methods must be taken into account, namely;

\begin{enumerate}
\item
The sum over the class of loops in series of \fig{fren}, using the Schwinger Dyson \eq{SDequation}.
The asymptote for the vertex of \eq{tpv1} contributes to the Schwinger Dyson sum, whereas
the asymptote of \eq{tpv2} gives a vanishing contribution for $n>1$. The Pomeron loop summation
generated by the Schwinger Dyson sum, provides the renormalized Pomeron intercept derived
in \eq{renormalizedPomeronintercept}.

\item
The sum over the symmetric class of loops in \fig{fsopl}, using the vertex of \eq{tpv2}.
This leads to the sum over even numbers of non interacting Pomerons, with renormalized Pomeron
vertices.
The symmetric nature of the loops in \fig{fsopl}, leads to even numbers of independent Pomerons. For example
taking all loop branches outside in \fig{fsopl} (b), leaves 4 independent Pomerons.
\end{enumerate}

The two asymptotic expressions for the triple Pomeron vertex in \eq{tpv},
 lead to the above two entirely different types of loop summation. 
It should be stressed, that the above two treatments
do not lead to the same result,
and the scattering amplitude requires taking into account both methods.
The non interacting Pomeron solution (2), originates from the loops in \fig{fsopl}
 stretching in rapidity space until they fill up
 the gap between the projectile and target, and therefore become independent Pomeron exchanges.
 This phenomena does not occur for the loops
 in series shown in \fig{fren}, since the loop cannot become non interacting Pomerons,
 when there is more
 than one loop in series. The only class of diagram which can yield non interacting Pomerons, is the special class of
 loops shown in \fig{fsopl} (a), and the only asymptote for the vertex which
 can yield non  interacting Pomerons is \eq{tpv2}.
  Hence, the non interacting Pomeron solution requires a separate treatment from
 the Schwinger Dyson equation.\FIGURE[h]{\begin{minipage}{180mm}
\centerline{\epsfig{file=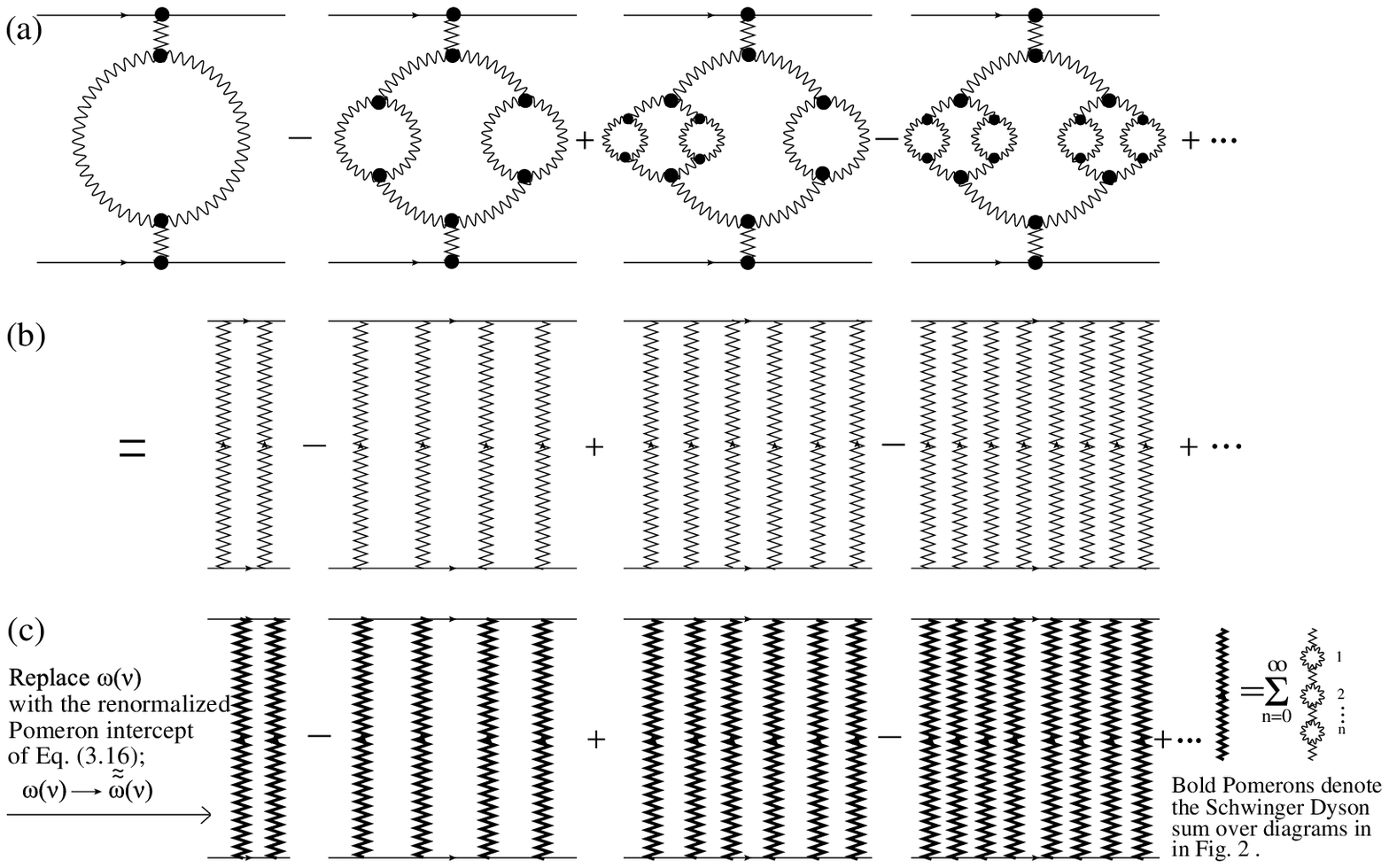,width=180mm}}
\end{minipage}
\caption{ The process which leads to the formula of \eq{sumoverPomeronloops}.
Summing over the symmetric class of loops in (a), leads to the sum over even numbers of non interacting Pomerons
shown in (b). Then replace the bare Pomerons with bold Pomerons, where bold Pomerons label the Schwinger Dyson sum over the class of loops in series shown in
\fig{fren}. The Schwinger Dyson sum renormalizes the Pomeron, by replacing the Pomeron intercept
with the renormalized Pomeron intercept $\widetilde{\tom}\Lb\nu\Rb$ of \eq{renormalizedPomeronintercept}. This leads to the sum over even numbers
of independent bold Pomerons in (c), that are renormalized in the framework of the Schwinger Dyson equation. }\label{fspl} }

The result of   \eq{sumoverPomeronloops} includes both types of summation over Pomeron loops.
\fig{fspl} shows a picture for the derivation of \eq{sumoverPomeronloops}.
It is instructive to first consider the class of loops in \fig{fspl} (a).
As explained in \sec{sNIP}, the summation over the class of loops in \fig{fspl} (a)
is equivalent to the sum over diagrams with an even number of non interacting Pomerons, shown in
\fig{fspl} (b). Next, replace the bare Pomerons with Pomerons which are renormalized by the Schwinger Dyson sum.
This means that for each of the non interacting Pomerons in \fig{fspl} (b), replace
it with the sum over loops in series shown in \fig{fren}, generated by the Schwinger Dyson equation.
This is achieved by replacing the bare Pomeron intercept $\omega\Lb\nu\Rb$, with the renormalized intercept $\widetilde{\tom}\Lb\nu\Rb$ derived in \eq{renormalizedPomeronintercept}.
This leads to the diagram of \fig{fspl} (c), where the bold Pomerons label the above described renormalized Pomerons with the intercept $\widetilde{\tom}\Lb\nu\Rb$.\\

The end result of \fig{fspl} (c) is a true description of \eq{sumoverPomeronloops}, namely the sum over diagrams
with an even number of independent Pomeron exchanges, which are renormalized in the context of the Schwinger Dyson
 equation. The formula of \eq{sumoverPomeronloops} does not include the bare scattering amplitude of \fig{fBFKL}
  given by \eq{A01}. The complete scattering amplitude which includes the loop corrections shown in \fig{fspl}
   is found by adding to \eq{sumoverPomeronloops}, the bare scattering amplitude of \eq{A01}. This would lead
    to a divergent result which violates unitarity. The only remedy for this problem, is to repeat the procedure
     shown in \fig{fspl}, for the non-symmetric class of diagrams shown in \fig{fsoplodd}.
Although the diagrams of \fig{fsoplodd} have been included in the Schwinger Dyson sum,
 this was performed using
 the vertex of \eq{tpv1}. The approach here, is instead to use the vertex of \eq{tpv2} to sum over the non
 symmetric diagrams in \fig{fsoplodd}. This
 will lead to the sum over odd numbers of non interacting Pomerons, that do not contribute to the Schwinger Dyson sum.

.\FIGURE[h]{\begin{minipage}{70mm}
\centerline{\epsfig{file=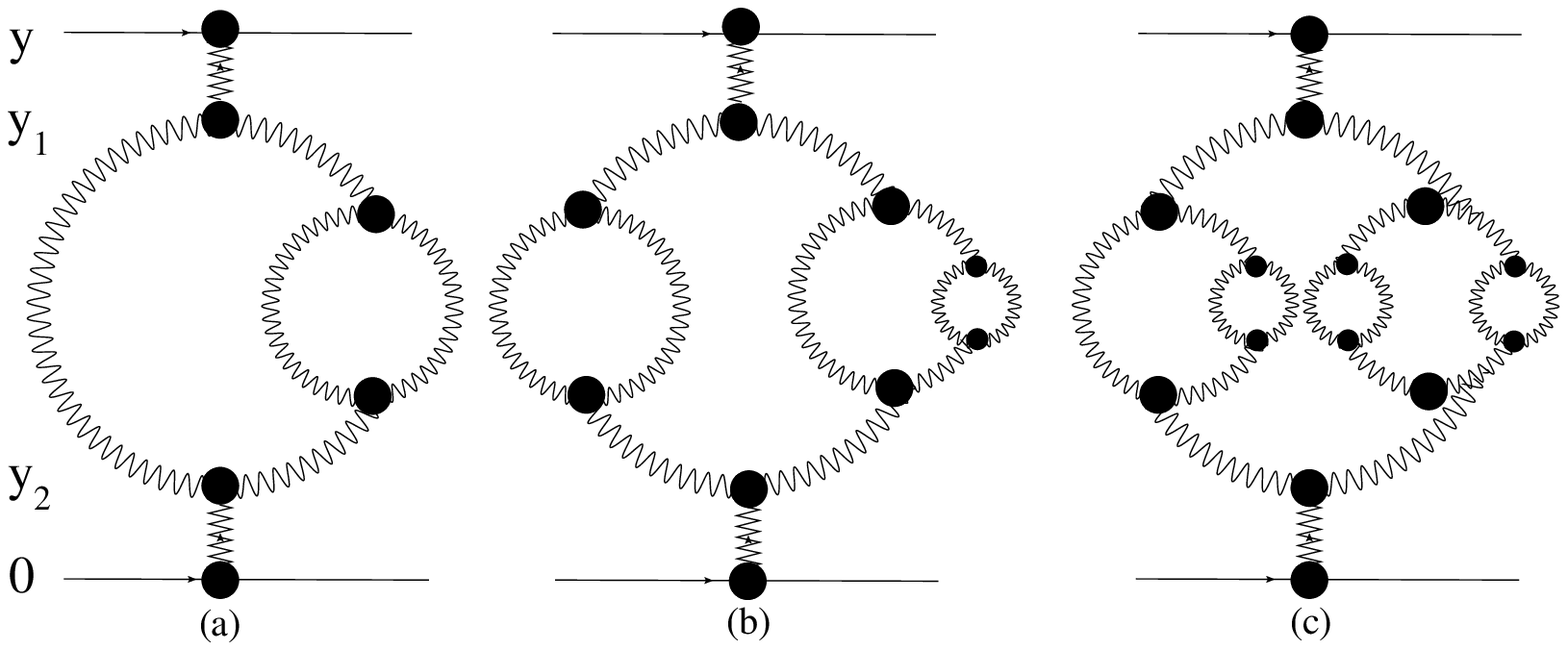,width=70mm}}
\end{minipage}
\caption{Examples of non symmetric Pomeron loop diagrams.} \label{fsoplodd} }

This stems from
the non symmetric nature of the loops in \fig{fsoplodd}, where for example
taking all branches of the loops in \fig{fsoplodd} (a) outside, reduces the diagram
to 3 independent Pomeron exchanges.
Following the same strategy described above, the independent Pomerons are
renormalized by the Schwinger Dyson sum, by replacing the Pomeron intercept with the renormalized intercept found in
\eq{renormalizedPomeronintercept}. Overall this yields the sum over odd numbers of renormalized non interacting Pomerons.\\

 Finally, adding the sum over odd numbers,
 to the sum over even numbers of renormalized Pomerons already derived in \eq{sumoverPomeronloops},
leads to the expression which takes the following form;

\bea
\sum^\infty_{n=0}A_{(n)}\Lb y|\mbox{symmetric + non symmetric diagrams}\Rb=\f{c\, e^{\widetilde{\tom}\Lb 0\Rb y}}{1+d  \,e^{\widetilde{\tom}\Lb 0\Rb y}}\lab{symmnonsymm}\eea

where $c$ and $d$ contain all the other terms which are part of the scattering amplitude. The key property of
 \eq{symmnonsymm}, is that it includes the basic amplitude of \fig{fBFKL}, and it preserves unitarity  generating a similar curve
to the one in \fig{fsed2}. \eq{symmnonsymm} is the pp elastic scattering amplitude, including the full set of Pomeron loop corrections.  Unfortunately, we have not yet been able to 
calculate the class of non symmetric diagrams shown in \fig{fsoplodd}, however this work is in progress.
In light of this discussion, the prospects for arriving at an expression which preserves unitarity, and includes symmetric and non symmetric loop diagrams, are hopeful.\\

\FIGURE[h]{\begin{minipage}{70mm}
\centerline{\epsfig{file=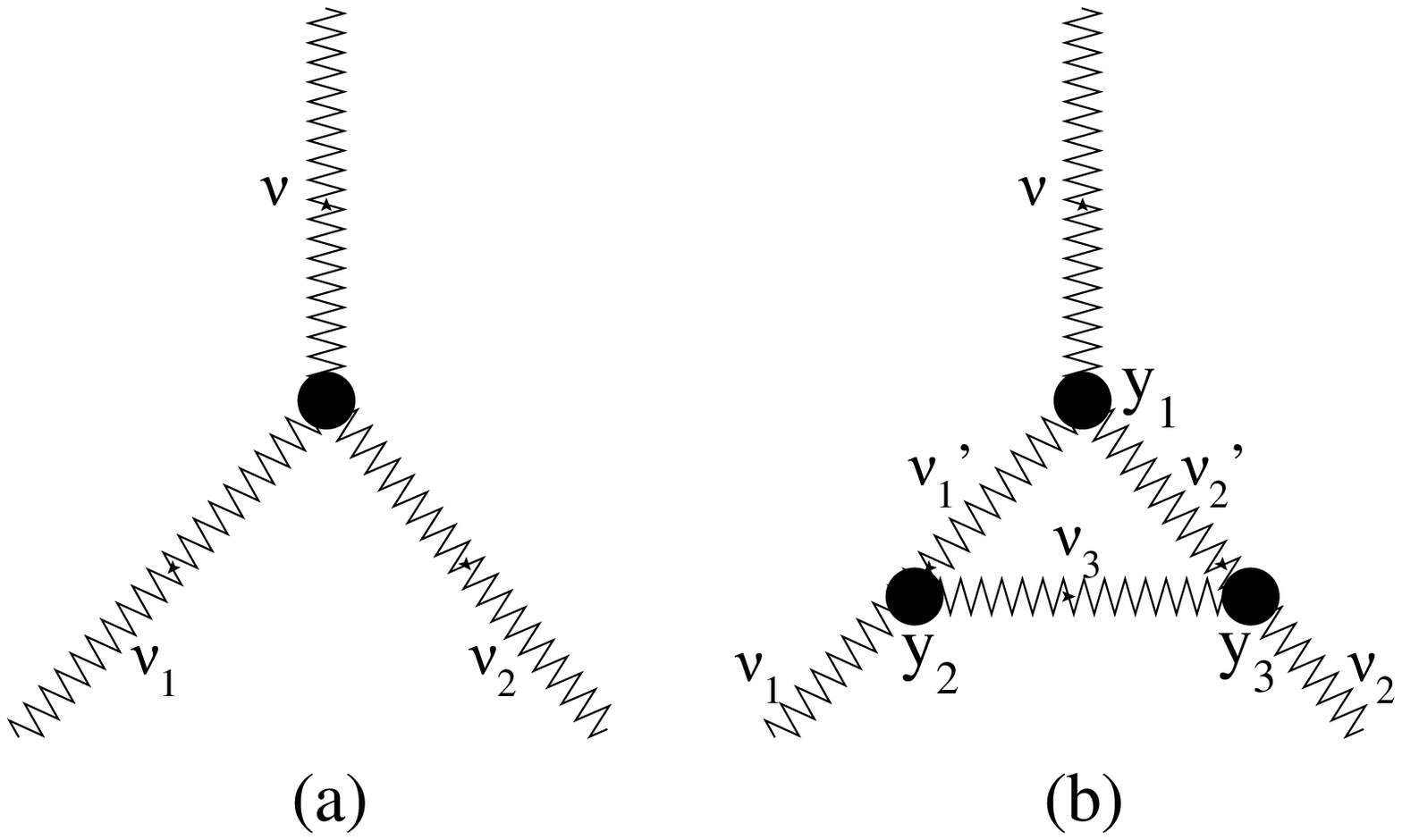,width=70mm}}
\end{minipage}
\caption{ Diagram (a) shows the lowest order triple Pomeron vertex, and diagram (b) shows the first order correction to the vertex.  } \label{fvertex} }In the calculations performed in this paper,  the diagrams which contribute to the vertex in the framework of the
Schwinger-Dyson equation, were not taken into account. Only the lowest order vertex shown in \fig{fvertex} (a), was included in the above performed calculations. 
To illustrate one example,
 the first order correction to the vertex is shown in \fig{fvertex} (b). The full triple Pomeron vertex which includes the complete set of corrections is
described by the Schwinger-Dyson equation for the vertex;

\vspace{0.8cm}

\begin{subequations}\lab{SDvertex}
\bea
&&\widetilde{\Ga}\Lb\nu,\nu_1,\nu_2| y_1\Rb=\Ga_{(0)}\Lb\nu,\nu_1,\nu_2\Rb\label{SDvertex1}\\\nn\\
&&-\int^{y_1}_0\!\!dy_2\int^{y_1}_0\!\!dy_3
\int^\infty_{-\infty}\!\!\!\!\mathcal{D}\nu_1^{\,\prime}\int^\infty_{-\infty}\!\!\!\!\mathcal{D}\nu_2^{\,\prime}
 \int^\infty_{-\infty}\!\!\!\!\mathcal{D}\nu_3\,\left\{e^{\omega\Lb\nu_1^{\,\prime}\Rb\,y_{12}}e^{\omega\Lb\nu_2^{\,\prime}\Rb\,y_{13}}
 e^{\omega\Lb\nu_3\Rb\,y_{23}}\right.\nn\\
 &&\left.\times
 \widetilde{\Ga}\Lb\nu,\nu_1^{\,\prime},\nu_2^{\,\prime}|y_1\Rb
\widetilde{\Ga}\Lb\nu_3,\nu_1,\nu_1^{\,\prime}|y_2\Rb\widetilde{\Ga}\Lb\nu_3,\nu_2,\nu_2^{\,\prime}|y_3\Rb\right\};\hspace{1cm}(y_{ij}=y_i-y_j);\nn\\
\nn\\
\mbox{where}\hspace{0.5cm}&&\int^\infty_{-\infty}\!\!\mathcal{D}\nu=\int^\infty_{-\infty}\!\!d\nu h\Lb\nu\Rb g\Lb\nu\Rb\,.\label{SDvertex2}
\eea
\end{subequations}

where $\Ga_{(0)}$ is the lowest order vertex shown in \fig{fvertex} (a), and $\widetilde{\Ga}$ is the full vertex which includes the complete set of vertex corrections  described by the
Schwinger-Dyson equation.
The Schwinger-Dyson \eq{SDvertex} for the vertex is much more complicated in comparison to the Schwinger-Dyson \eq{SDequationclear} for the
Pomeron Green function.
 The author acknowledges, that the set of corrections to the vertex, are also required for the formula for the scattering amplitude which includes all possible
corrections. This problem is very challenging owing to the complexity of the non-closed equation for the vertex of \eq{SDvertex}, but nevertheless attempts to solve this problem are in progress. 
The two expressions used for the lowest order triple Pomeron vertex in \eq{tpv}, indicate that the vertex is less than unity. Thus,
although
the triple Pomeron vertex was not taken into account in the framework of the Schwinger-Dyson \eq{SDvertex}, since the vertex is less than $1$, corrections to the vertex are expected to
give a small contribution.

In summary, the main achievements of this article include the following;

\begin{enumerate}
\item
A closed solution to the Schwinger-Dyson equation in perturbative QCD, which generates the summation over the full set of Pomeron loops, leading to the renormalized
Pomeron intercept.

\item
A closed expression for the summation over a special class of Pomeron loop diagrams, equivalent to non interacting Pomerons which preserves unitarity.
\end{enumerate}

Both of these achievements are original, and provide a strong foundation for calculating the scattering amplitude in perturbative QCD.
The remaining corrections which are required for the scattering amplitude, include the non symmetric loop diagrams of
\fig{fsoplodd} and the corrections to the vertex in the formalism of the Schwinger-Dyson equation. The calculation of these additional required 
corrections, is the next challenging problem to be solved.

We would like to
thank G. Milhano  for their careful reading and helpful advice in writing this paper.  We would
also like to thank S.Abereu, L. Apolin$\acute{a}$rio, M. Braun, J. Dias De Deus and E. Levin
for fruitful discussions on the subject.
This research was supported by the Funda\c{c}$\tilde{a}$o para ci$\acute{e}$ncia e a tecnologia (FCT), and CENTRA - Instituto Superior T$\acute{e}$cnico (IST), Lisbon.

\end{document}